# Rheological characterization of a thixotropic semisolid slurry by means of numerical simulations of squeeze flow experiments

Georgios C. Florides, Georgios C. Georgiou
Department of Mathematics and Statistics, Univerty of Cyprus, PO Box 20537, 1678 Nicosia, Cyprus
gcflorides@eac.com.cy, georgios@ucy.ac.cy

Michael Modigell
Aachener Verfahrenstechnik, RWTH Aachen University, Turmstraße 46, 52056 Aachen, Germany
michael.modigell@avt.rwth-aachen.de

Eugenio José Zoqui
School of Mechanical Engineering, University of Campinas, Brazil
zoqui@unicamp.br

**Abstract**

We propose a methodology for the rheological characterization of a semisolid metal slurry using experimental squeeze flow data. The slurry is modeled as a structural thixotropic viscoplastic material, obeying the regularized Herschel-Bulkley constitutive equation. All rheological parameters are assumed to vary with the structure parameter that is governed by a first-order kinetics accounting for the material structure breakdown and build-up. The squeeze flow is simulated using finite elements in a Lagrangian framework. The evolution of the sample height has been studied for wide ranges of the Bingham and Reynolds numbers, the power-law exponent as well as the kinetics parameters of the structure parameter. Systematic comparisons have been carried out with available experimental data on a semisolid aluminium alloy (A356), where the sample is compressed from its topside under a specified strain of 80% at a temperature of 582 ºC while the bottom side remains fixed. Excellent agreement with the experimental data could be achieved provided that at the initial instances (up to 0.01s) of the experiment the applied load is much higher than the nominal experimental load and that the yield stress and the power-law exponent vary linearly with the structure parameter. The first assumption implies that a different model, such as an elastoviscoplastic one, needs to be employed during the initial stages of the experiment. As for the second one, the evolution of the sample height can be reproduced allowing the yield stress to vary from 0 (no structure) to a maximum nominal value (full structure) and the power-law exponent from 0.2 to 1.4, i.e., from the shear-thinning to the shear-thickening regime. These variations are consistent with the internal microstructure variation pattern known to be exhibited by semisolid slurries.

# 1. Introduction

Semisolid metal (SSM) processing is an established technology used for the production of complex near-net-shaped components with high quality and durability characteristics, used mainly in the automotive and electronics industries [1]. In contrast to conventional metal forming technologies, the prime material is a dense suspension of specially prepared solid particles with globular or rosette-type shape lying in a state between the solidus and liquidus limits. The average solid-liquid interchange depends on the temperature. The process is also called thixoforming, where the material shows a solid behavior under undisturbed conditions and flows when sufficient shear load is applied [2-4]. Unlike fully liquid metals, which show Newtonian behavior, semi-solid suspensions exhibit viscoplastic flow behavior, due to the interaction of the solid-liquid matrix, and thixotropic characteristics, i.e., and history-dependent material-parameters [1]. The entire phenomenon appears to be irreversible as the continuous solid skeleton breaks down under constant shearing [5,6]. Semisolid processing of metallic alloys and composites exploits the thixotropic behavior of these materials in the semisolid state [7]; the material thins when sheared and thickens again when shearing ceases [8]. Due to the high processing temperatures, estimating the rheological properties of semi-solid metal slurries is not an easy task. Thus, different engineering approaches have been proposed to that end [9]. It should be noted that it has been reported that semisolid metal slurries may exhibit not only shear thinning, but also shear thickening [7,8].

Comprehensive reviews of thixotropy have been carried out by Barnes [10], Mewis and Wagner [11] and Larson [12]. Larson and Wei [13] have also recently reviewed thixotropy and its rheological modeling. Thixotropy is usually exhibited by colloidal suspensions, clay suspensions, semisolid metal suspensions, cement pastes, paints, food products and biological fluids [10]. These materials, which are encountered in a wide range of industrial applications, are also known to be viscoplastic [14], and are thus referred to as time-dependent yield-stress materials [15]. The difficulties in measuring the yield stress of thixotropic materials has been discussed by Møller et al. [16].

Thixotropy is a phenomenon associated with the material microstructure, which breaks down or builds up depending on the shearing history [13]. The viscosity is gradually reduced when shear is applied, increases with time when shearing is stopped, and can recover its initial state after a long enough resting [17,18]. Hence, the complex rheological behavior of thixotropic materials is a consequence of the competition between rejuvenation (destructuring) and aging (restructuring or build-up). Thixotropy is usually modeled using structural kinetics and assuming that the viscosity and other rheological parameters depend on a dimensionless time-dependent structure parameter that follows a kinetics equation, in analogy with chemical reaction kinetics. The values of the structure parameter range from 0 (unstructured) to 1 (fully structured) [19]. As pointed out by Varchanis et al. [20], the structure parameter may represent different things, such as the number of welded bonds in semi-solid slurries, integrity of particles aggregates in suspensions, entanglements and integrity of polymeric chains in gels, etc.

The squeeze or compression flow is a transient viscometric test used to characterize viscoplastic materials [14]. Hot compression tests are often used to characterize the rheological behavior of semisolid metals, especially those of a solid fraction higher than 0.5 [1], which is essential in thixoforming processes [21]. In these viscometric tests, a fixed amount of material is squeezed under constant force or velocity and information about its rheological behavior is deduced from the relation between the force and the displacement of the sample. Modigell et al. [1] note that the viscosity at a desired shear rate is calculated under the assumption that the material is Newtonian, i.e., the calculated viscosities are apparent only.

The numerical simulation of the squeeze flow of viscoplastic materials has been the subject of many experimental, theoretical and numerical works in the past few decades. These have been recently reviewed by Muravleva [22], who also employed the accelerated augmented Lagrangian method along with finite differences to study the squeeze flow under constant velocity of Bingham, Herscchel-Bulkley and Casson fluids in the presence of wall slip.

Some years ago, we have employed a mixed-Galerkin finite element method and the Papanastasiou regularization to numerically solve the compression flow of a Bingham plastic in Lagrangian coordinates [23]. Numerical simulations were carried out considering both constant load and constant velocity and the effects of the Reynolds and Bingham number on the flow and the shape evolution were investigated. Emphasis was also given on the development of the yielded/unyielded regions during the compression experiment. This work was subsequently extended [24] to account for thixotropic effects, allowing the yield stress to vary linearly with a structure parameter, assumed to follow a first-order rate equation. We studied in particular the development of the yielded/unyielded regions in relation to material structural changes and reported that under constant force the structure may be destroyed at the early stages of the compression, but it re-builds steadily at a later time till the cessation of the experiment [24]. In these works, no attempts for comparisons with experimental observations and the rheological characterization of the semi-solid slurries were made. As pointed out by de Souza Mendes and Thompson [15], the latter task is quite challenging due to the complex behavior of time-dependent yield stress materials.

In this paper, we propose a methodology for the rheological characterization of semisolid metal slurries from squeeze flow experiments using numerical simulations. To this end, we extend the structural thixotropic model used in Ref. [24] by employing the regularized Herschel-Bulkley instead of the Bingham-plastic constitutive equation and by allowing all the rheological parameters (yield stress, power-low exponent, and consistency index) to vary with the structure parameter. The proposed methodology is tested against experimental data on a semisolid aluminum A356 alloy obtained applying a nominal load on the top of a cylindrical sample. We carry out systematic numerical simulations of the compression test imposing the experimental load on the semisolid sample and varying material and flow parameters. In order to reproduce the observed initial fast compression rate, we have also investigated alternative load distributions only in the very initial stages of the experiment. These simulations showed that it is possible to reproduce the experiments provided that: (a) the yield stress and the power-law exponent are assumed to vary linearly with the structure parameter; and (b) during the very early moments of the experiment, i.e., for the first 0.01s, the applied load is much higher than the nominal applied load.

The rest of the paper is organized as follows. In Section 2, the experimental set-up and data are briefly presented. The formulation of the squeeze flow problem and the structural thixotropic constitutive equation are discussed in Section 3, where the numerical method is also described. The numerical results are presented in Section 4. First, systematic numerical simulations when using the experimental load distribution are discussed. These results demonstrate that the evolution of the sample height can only be achieved by using a modified load distribution, where a load much higher than the nominal one is applied only for the first 0.01s of the experiment. Subsequently, it is also demonstrated that the squeeze flow experiment can be accurately simulated by allowing the power-law exponent to increase with the structure parameter. The values of the material parameters of the semisolid material at the experimental conditions have also been obtained by using the optimal parameters of the dimensionless values describing the squeeze flow.

## 2. Experiments

Squeeze flow data have been obtained on an A356 alloy in the semisolid state at a specified strain of 80%. The cylindrical sample was compressed from the topside. For the preparation of the A356 sample a cooling die was used under mechanical vibration of 10 Hz. The sample of height $H_0^* = 30.5 \text{ mm}$ and radius $R^* = 15 \text{ mm}$ was heated up to the semisolid temperature (582 ºC) under an average heating rate of 27.5 ºC/min. The total time of heating was 20.4 min. More details about the material composition, the experimental set-up, and the microstructure analysis are provided in Refs. [25,26].

Figure 1 shows the applied load and the evolution of the sample's height. Roughly speaking the load increases linearly initially for 0.25 seconds till a final specified (nominal) force, $F_\infty^* \approx 9$ kN, is reached (Fig. 1a). It should be noted that the total duration of the squeeze test is just 0.9 s. The sample height decreases from 30.5 mm to 6.23 mm. It is clear that in the initial stage of the experiment the sample height is reduced linearly with time, which implies that the compression velocity is constant initially. It then decreases with time and diminishes at the final stage of the compression test, where the sample height is essentially constant (Fig. 1b).

## 3. Governing equations and numerical method

Let $\boldsymbol{\tau}^*$ be the viscous stress tensor, and $\mathbf{D}^*$ be the symmetric part of the velocity-gradient tensor defined through

$$\mathbf{D}^* \equiv \frac{1}{2}\left[\nabla \mathbf{u}^* + (\nabla \mathbf{u}^*)^T\right] \quad (1)$$

where $\mathbf{u}^*$ is the velocity vector and the superscript $T$ denotes the transpose. The constitutive equation for a Herschel-Bulkley fluid can then be written as follows [27,28]:

$$\begin{cases} \mathbf{D}^* = \mathbf{0}, & \tau^* \leq \tau_y^* \\ \boldsymbol{\tau}^* = 2\left(\dfrac{\tau_y^*}{\dot{\gamma}^*} + k^* \dot{\gamma}^{*n-1}\right) \mathbf{D}^*, & \tau^* > \tau_y^* \end{cases} \quad (2)$$

where $\dot{\gamma}^* \equiv \sqrt{2 \operatorname{tr} \mathbf{D}^{*2}}$ and $\tau^* \equiv \sqrt{\operatorname{tr} \boldsymbol{\tau}^{*2}/2}$ are the magnitudes of $2\mathbf{D}^*$ and $\boldsymbol{\tau}^*$, respectively. Note that $\operatorname{tr}\mathbf{M}$ stands for the trace of any matrix $\mathbf{M}$. The Herschel-Bulkley fluid involves three material parameters: the yield stress, $\tau_y^*$, below which the fluid is at rest or moves as a rigid plug; the consistency index, $k^*$, and the power-law exponent or flow index, $n$. It should be noted that symbols with stars denote dimensional quantities.

The two-branch form and the resulting non-differentiability of constitutive equation (2) causes severe difficulties in simulating Herschel-Bulkley flows, since the regions where the material is yielded ($\tau^* > \tau_y^*$) or unyielded ($\tau^* \leq \tau_y^*$) need to be determined. To avoid this, we use a regularized version of the constitutive equation that holds uniformly in the entire flow domain (in both yielded and unyielded regions). More specifically, we use Papanastasiou's regularization [29]:

$$\boldsymbol{\tau}^* = 2\left\{\frac{\tau_y^*\left[1-\exp(-m^*\dot{\gamma}^*)\right]}{\dot{\gamma}^*} + k^*\dot{\gamma}^{*n-1}\right\} \mathbf{D}^* \quad (3)$$

where $m^*$ is the regularization parameter, which has units of time. This parameter needs to be sufficiently high so that Eq. (3) approximates satisfactorily the original Herschel-Bulkley constitutive equation. The alternative to the regularization approach is the use of the augmented Langrangian method. The advantages and limitations of both approaches are reviewed in Refs. [30,31].

To account for the thixotropic properties of the semisolid material of interest a structural model is employed. The rheological parameters are assumed to be functions of a time-dependent structure parameter $\lambda(t^*)$, whose values range from zero (no structure) to unity (complete structure) [11,17,18]. The original structural model with all the material parameters of the Herschel-Bulkley fluid assumed to be functions of $\lambda(t^*)$ was proposed by Houska [32]:

$$\begin{cases} \mathbf{D}^* = \mathbf{0}, & \tau^* \leq \tau_y^*(\lambda) \\ \boldsymbol{\tau}^* = 2\left(\dfrac{\tau_y^*(\lambda)}{\dot{\gamma}^*} + k^*(\lambda)\dot{\gamma}^{*n(\lambda)-1}\right)\mathbf{D}^*, & \tau^* > \tau_y^*(\lambda) \end{cases} \quad (4)$$

In the present work, we assume that the yield stress and the power-law exponent vary linearly with the structure parameter:

$$\tau_y^* = \tau_0^* \lambda(t^*) \quad (5)$$

and

$$n(\lambda(t^*)) = n_{\min} + (n_{\max} - n_{\min})\lambda(t^*) \quad (6)$$

where $\tau_0^*$ is the maximum yield stress (corresponding to the fully-structured material) and $n_{\min}$ and $n_{\max}$ are the minimum and maximum values of the exponent, respectively. For the sake of simplicity, the numerical value of the consistency index is assumed to remain constant. However, it should be noted that $k^*$ also varies with the structure parameter, since its units depend on the power-law exponent ($\left[k^*\right] = \text{Kg/m/s}^{2-n(\lambda)}$). The above assumptions imply that when structure is completely destroyed, the material is not viscoplastic (the yield stress vanishes) and behaves as a power-law fluid with a flow index equal to $n_{\min}$. According to Larson and Wei [13], thixotropic/viscoplastic models, like the one used here, are known as ideal thixotropic models (there is no elasticity or viscoelasticity in the model); see also the interesting discussion of de Souza Mendes and Thompson [15] about the dependence of all material parameters of the Herschel-Bulkley model on the structure parameter. Finally, it is assumed that $\lambda(t^*)$ follows the following first-order kinetics:

$$\frac{D\lambda}{Dt^*} = a^*(1-\lambda) - b\lambda\dot{\gamma}^* e^{c^*\dot{\gamma}^*} \quad (7)$$

where $D\cdot/Dt^*$ is the material derivative, $a^*$ (reciprocal time units) is the recovery parameter, and $b$ (dimensionless) and $c^*$ (in time units) are positive breakdown parameters. The two terms of the RHS correspond to the build-up (recovery or restructuring) and breakdown or destructuring rates of the structure parameter [11,17,18]. It can be observed that the rate of recovery is proportional to the fraction $(1-\lambda)$ of the broken links, whereas the break down rate depends not only on the fraction $\lambda$ of the unbroken links, but also on the deformation rate $\dot{\gamma}^*$. For alternative rate equations, the reader is referred to Refs. [17,18]. In summary, the proposed thixotropic model involves seven material parameters, i.e., $\tau_0^*, n_{\min}, n_{\max}, k^*, a^*, b$ and $c^*$; as mentioned above the regularization parameter $m^*$ is taking a sufficiently large value.

To our knowledge, the first work where the Herschel-Bulkley constitutive equation was modified to include a structure parameter accounting for time-dependent effects was that of Tiu and Boger [33], who rheologically characterized mayonnaise samples. They assumed linear variations of the yield stress and the consistency index with the structure parameter. In their study on unclassified tailing pastes, Yang et al. [34] also considered the Herschel-Bulkley model assuming that only the yield stress changes with the structure parameter. Toorman [19] employed the Bingham model with the yield stress and the plastic viscosity being functions of the structure parameters to model the thixotropic behavior of dense clay suspensions. Malmir et al. [35] modeled the rheological behavior of mature fine tailings using the Bingham model and assuming that both the yield stress and the plastic viscosity vary linearly with the structure parameter. Similarly, Guo et al. [36] employed the Herschel-Bulkley constitutive equation and structural kinetics to model the transient flow of superfine-tailings cemented paste backfill, assuming that the yield stress and the consistency index also vary with the structure parameter, while the power-law exponent remains constant.

The geometry and the boundary conditions are illustrated in Fig. 2. The cylindrical sample of initial height $H_0^*$ and radius $R^*$ is confined between two parallel plates. The upper plate moves downwards subject to a specified applied load

$$F^*(t^*) = \int_S \left(-p^* \mathbf{I} + \boldsymbol{\tau}^*\right) \cdot \mathbf{e}_z dS^* \tag{8}$$

while the lower plate remains fixed. $F^*(t^*)$ initially increases with time and reaches a nominal load $F_\infty^*$ at a finite time. The bottom plate is fixed and thus both velocity components vanish there (no slip and no penetration). Symmetry boundary conditions are applied along the symmetry axis, while on the free boundary of the sample surface tension is assumed to be negligible.

In order to dedimensionalize the flow problem, lengths are scaled by $H_0^*$, the velocity vector by $(F_\infty^*/(k^* H_0^{*2}))^{1/n_{\max}} H_0^*$, time by $(k^* H_0^{*2}/F_\infty^*)^{1/n_{\max}}$, the load by $F_\infty^*$, and the pressure and stress components by $F_\infty^*/H_0^{*2}$. The stars are simply dropped in order to denote the dimensionless counterparts of all variables. The continuity and momentum equations then become:

$$\nabla \cdot \mathbf{u} = 0 \tag{9}$$

and

$$Re\left(\frac{\partial \mathbf{u}}{\partial t} + \mathbf{u} \cdot \nabla \mathbf{u}\right) = -\nabla p + \nabla \cdot \boldsymbol{\tau} \tag{10}$$

where the Reynolds number is defined by:

$$Re \equiv \frac{\rho^* F_\infty^{*2/n_{\max}-1}}{k^{*2/n_{\max}} H_0^{*4(1/n_{\max}-1)}} \tag{11}$$

$\rho^*$ being the constant density of the material.

Given our assumption that the power-law exponent varies with time, special attention needs to be paid to the non-dimensionalization of the constitutive equation. With the above scalings, one gets:

$$\boldsymbol{\tau} = 2\left\{\frac{Bn\lambda[1-\exp(-M\dot{\gamma})]}{\dot{\gamma}} + K(\lambda)\dot{\gamma}^{n(\lambda)-1}\right\}\mathbf{D} \tag{12}$$

where

$$Bn \equiv \frac{\tau_0^* H_0^{*2}}{F_\infty^*} \tag{13}$$

is the Bingham number,

$$M \equiv m^* \left( \frac{F_\infty^*}{k^* H_0^{*2}} \right)^{1/n_{max}} \tag{14}$$

is the regularization number, and

$$K(\lambda) \equiv \left( \frac{F_\infty^*}{k^* H_0^{*2}} \right)^{n(\lambda)/n_{max} - 1} \tag{15}$$

is the dimensionless consistency index. Note that $K = 1$ when the power-law exponent is constant ($n_{min} = n_{max} = n$). In the general case, however, the value of $K$ should be locally adjusted so that the relevant term remains dimensionless, since the units of the consistency index vary with the power-law exponent and the dedimensionalization, of the shear rate is based on its maximum value $n_{max}$.

Finally, the dimensionless version of Eq. (7) is

$$\frac{D\lambda}{Dt} = a(1-\lambda) - b\lambda \dot{\gamma} e^{c\dot{\gamma}} \tag{16}$$

where

$$a \equiv a^* \left( \frac{k^* H_0^{*2}}{F_\infty^*} \right)^{1/n_{max}} \tag{17}$$

and

$$c \equiv c^* \left( \frac{k^* H_0^{*2}}{F_\infty^*} \right)^{1/n_{max}} \tag{18}$$

The dimensionless experimental load applied on the top side of the compression samples is simply $F \equiv F^*(t^*)/F_\infty^*$.

For the numerical simulations, we use Lagrangian coordinates and resort to the Arbitrary Eulerian-Lagrangian formulation [24], with which the shape of the free surface is obtained directly, and employ a mixed Galerkin finite element method. The pressure and velocity fields are approximated by means of bilinear and biquadratic basis functions, respectively. The Newton method is used to solving the non-linear algebraic systems at each time step. Also, remeshing is applied depending on the sample deformation by means of a Laplace-type discretization algorithm. For more information about the numerical method the reader is referred to Refs. [23,24].

## 4. Numerical results

Based on the convergence studies in our previous works [23,24], in all numerical simulations we employed a mesh consisting of $20 \times 20$ elements and set the time step to $\Delta t^* = 0.001$ s and the regularization parameter to $M = 300$. The latter value appears to be sufficiently high so that the regularized constitutive equation approximates adequately the ideal Bingham-plastic equation.

We have carried numerical simulations of the squeeze flow experiment using two different load distributions:
(i) The imposed load distribution $F^*(t^*)$ follows closely the experimental distribution and eventually attains the nominal experimental load $F^*_\infty = 9$ kN, as illustrated in Fig. 3a.
(ii) The imposed load is initially set to a constant value $F^*_0 \geq F^*_\infty$ for a short period of time $\Delta t^*$, which is a small fraction of the duration of the actual experiment (approximately 0.9 s); after this short period the load is set to the nominal experimental load $F^*_\infty$ (Fig. 3b). Hence, in this case, the imposed load is

$$F^*(t^*) = \begin{cases} F^*_0, & 0 \leq t^* \leq \Delta t^* \\ F^*_\infty, & t^* > \Delta t^* \end{cases} \tag{19}$$

In order to facilitate the discussion of the results, the load ratio is defined as follows

$$f_r = \frac{F^*_0}{F^*_\infty} \tag{20}$$

Hence, Eq. (19) also takes the form

$$\frac{F^*(t^*)}{F^*_\infty} = \begin{cases} f_r, & 0 \leq t^* \leq \Delta t^* \\ 1, & t^* > \Delta t^* \end{cases} \tag{21}$$

As discussed below, only with the latter distribution it has been possible to describe well the evolution of the height at the initial stages of the experiment. As pointed out by Coussot [14], deformations in the solid regime can play a critical role in transient flows.

The two sets of numerical experiments are discussed in the following two subsections. In subsection 4.1, the experimental load distribution is applied and the effects of the various flow and material parameters are systematically studied. It is concluded that the experiment cannot be reproduced satisfactorily when the nominal experimental load is used and the power-law index is constant. If, however, $n$ varies with the structure parameter, following Eq. (6), then the final stages of the experiment can be simulated, but still the reproduction of the initial stages of the experiment is not good: the reduction of the sample height is much slower than in the experiment. This suggests the use of a much higher load only at the initial stages of the experiment, as in Eq. (21). As discussed in subsection 4.2, using a modified load with $f_r = 10$ and allowing the power-law index to vary with the structure parameter, resulted in the accurate reproduction of the squeeze experiment.

### 4.1 Simulations with the experimental load distribution
We first investigated the effects of the Bingham and Reynolds numbers, when the load is the experimental one (Fig. 3a) and the power-law index is fixed, i.e., only the yield stress varies with the structure parameter following Eq. (5). Unless otherwise indicated, the dimensionless numbers governing the dynamics of the structure parameter in Eq. (16) were $a = b = 1$ and $c = 0.0001$. The effect of the Bingham number on the sample height evolution for $Re = 1$, 1.5 and 0.5 is illustrated in Figs. 4, 5, and 6, respectively. In Figs. 4 and 5, results for $n = 1.2$, 1.0, 0.8 and 0.6 are shown. Note that at low values of the Reynolds number, e.g., for $Re = 0.5$ (Fig. 6), the method diverges and no results have been obtained for $n = 0.6$ (the compression stops at the very initial steps of the simulation). It should be mentioned that the simulations for $n = 1$ (Bingham fluid) are in good agreement with the squeeze-flow simulations of Alexandrou et al. [24] under constant load.

It can be observed in Fig. 4, where $Re = 1$ and the Bingham number, *Bn,* was varied from 0.00025 to 0.002, that the rate of compression is very slow initially, since the experimental load distribution (Fig. 3a) at this stage is very low and increases linearly up to the nominal experimental load. The rate of compression, however, increases accordingly and as the *Bn* number increases it becomes faster in all cases. As shown in Fig.4(d), the final stage of the experiment, i.e., at the stage where the compression of the material becomes very slow, can be better described with low values of the Bingham number and the power-law exponent, *Bn*=0.0005 and *n*=0.6.

In Figs. 5 and 6 we examine further the effect of Bingham number, *Bn,* for Re=1.5 (Figure 5) and Re=0.5 (Fig. 6) using the same parameters as in Figure 4. As already mentioned, when $Re = 0.5$ and $n = 0.6$ the method diverges and now results were obtained. Comparing the results in Figs. 4d and 5d, where *Bn*=0.0005 and *n*=0.6, one observes that the results with *Re*=1.0 approach more closely the experimental data at the final stage of the experiments. The results with *Re*=0.5 (Fig. 6) clearly fail to reproduce satisfactorily the experimental curve, even at the final stage of the compression.

Guided by the simulations in Figs. 4-6, we chose the value of $Bn = 0.001$ and investigated the effect of the Reynolds number for different values of the power-law exponent. To our knowledge, this is the first time where the effect of the power-law exponent is investigated in squeeze-flow under constant load. Results for $n = 1.2$, 1.0, 0.8 and 0.6, are shown in Fig. 7. Note that the effect of the Reynolds number is not important at the initial and final stages of the squeeze experiment. Initially, the compression appears to be very slow, independently of the value of the Reynolds number, which is in complete disagreement with the experimental data. The curves for different values of the Reynolds numbers merge in the final stages of the compression experiment, where compression becomes very slow. For certain choices of the flow and material parameters, the numerical results coincide with the experimental curve; see, e.g., Fig. 7c, where $n = 0.8$. However, in the intermediate stage of the squeeze flow, the numerical rate of compression appears to become faster as the Reynolds number is reduced. Another useful observation is that the compression becomes faster and a smaller final sample height is obtained as the power-law exponent is reduced. Since structure is expected to be destroyed at the final stage of the compression, it is reasonable to assume that yield stress vanishes and the rheology of the material can better be described using a lower value of the power-law exponent.

Figure 8 illustrates the effect of the power-law exponent for three combinations of the Reynolds and Bingham numbers, i.e., $Re = 1, Bn = 0.001$ (Fig. 8a), $Re = 1.5, Bn = 0.001$ (Fig. 8b), and $Re = 1, Bn = 0.0015$ (Fig. 8c). In all cases, the numerical results are close to the experimental data only in the final stages of the compression. It is clear that it is not possible to achieve a satisfactory simulation of the initial stages of the compression under the assumption of a constant power-law exponent. Note also that the effect of the power law exponent becomes important at different stages of the compression experiment, as the structural state of the material changes continuously during compression. This observation naturally leads to the idea of allowing the power-law index to vary with the structure parameter during compression, that is to vary with the local solid-liquid state of the material.

Figure 9 shows results for $Re = 1$ and different ranges of the power-law exponent. Fig. 9a shows results for $Bn = 0.001$ and three different ranges of the power-law exponent, from 0.6 up to 1.4. It can be seen that when $n_{min}$ is reduced compression becomes faster and the evolution of the height in the final stages of the experiment is better described. We note in Fig. 9b, that reducing the minimum value of the power-law exponent to $n_{min} = 0.2$ and the Bingham number to $Bn = 0.00025$ the final stage of the compression is very accurately reproduced.

In Figs. 10 and 11, the combined effects of the Reynolds number and the value of $n_{min}$ for $Bn = 0.001$ and $Bn = 0.00025$, respectively are illustrated. As already pointed out, the effect of the Reynolds number is important only in the intermediate stage of the experiment, the duration of which grows as the Bingham number is reduced. By comparing Figs. 10 and 11, we observe that as the Reynolds number is increased the rate of compression initially becomes slower but then it increases resulting in a reduced final sample height. For the lower value of the Bingham number ( $Bn = 0.00025$ in Fig. 11), the compression curves for $Re = 1.5$, 1 and 0.5 become horizontal, which is also the case with the experiment. For the higher value $Bn = 0.001$ in Fig. 10, the compression curves for the three values of $Re$ appear to merge but they do not become horizontal as dictated by the experiments. From Figs. 10 and 11, it can also be deduced that in order to obtain the final experimental height for a lower Bingham number, we need to reduce the minimum value of the power-law exponent to $n_{min} = 0.2$ and set the value of the Reynolds number around unity.

In Fig. 12, these optimal values are selected and various combinations of the two thixotropic parameters, $a$ and $b$, are used in order to illustrate their effects when $Re = 1$, $Bn = 0.00025$, $n_{min} = 0.2$, and $n_{max} = 1.4$. As expected, the rate of compression becomes faster and the final sample height decreases when the build-up parameter $a$ is reduced. Reducing the break-down parameter $b$ leads to an opposite and more pronounced effect. The third thixotropy parameter $c$ was kept constant ($c=0.0001$) as this does not have a significant contribution in the rate of compression. Our experimentation with different values of $a$, $b$ and $c$ showed that with the optimal values $a = b = 0.95$, $c=0.0001$, $Re = 0.95$, $Bn = 0.000235$, $n_{min} = 0.2$, and $n_{max} = 1.4$, the final stage of the compression experiment is very accurately reproduced. It should be noted that the very small optimal value of the $c$, tells us that the contribution of the exponential term in the kinetics equation (16) is negligible. In other words, the kinetics equation used is essentially the following:

$$\frac{D\lambda}{Dt} = a(1-\lambda) - b\lambda\dot{\gamma} \qquad (22)$$

Our numerical experiments with the experimental load profile failed to reproduce the initial stage of the compression experiment for wide ranges of the flow parameters, predicting a rather slow compression rate so that the sample appears to remain uncompressed initially. This difficulty along with the fact that the resistance of the material particles due to bonding, dry friction and hydrodynamic forces needs to be overcome at the initial stage of the compression experiment [37,38] prompted the idea of experimenting with a modified load distribution that follows Eq. (21). The results are discussed below.

### 4.2 Simulations with the modified load distribution
Our numerical experiments with the optimal material parameters for different values of the two load parameters that appear in the modified load distribution load of Eq. (21), i.e., the load ratio $f_r$ and the duration $\Delta t^*$ of the initial high load, showed that the optimal values are $f_r = 10$ (the initial load is ten times the nominal maximum experimental load $F_\infty^* = 9$ kN) and $\Delta t^* = 0.01$ s. Note that the duration of the experiment is just 0.9 s and that the 'optimal' values of all the other parameters were used: $Re = 0.95$, $Bn = 0.00025$, $n_{min} = 0.2$, $n_{max} = 1.4$, $a = b = 0.95$, and $c = 0.0001$. With these values the agreement between the model and the experiment is excellent; see Fig. 13.

Representative results illustrating the effects of the parameters of the modified load distribution are shown in Fig. 14. These effects are more important in the early stages and become less pronounced in the last stage of the compression. In Fig14a, the sample heights for $\Delta t^* = 0.01$ s and various values of $f_r$ are plotted. Increasing the value of $f_r$ accelerates in initial compression rate and reduces the final sample height. It is clear that this is necessary in order to obtain qualitative and quantitative agreement with the experiments. Increasing the value of $\Delta t^*$ when the ratio $f_r$ is fixed has a similar effect on the compression height; see Fig. 14b.

The optimal values of the flow and material parameters were further scrutinized. Figure 15 shows how the numerical simulation is affected when choosing values of the Bingham number (Fig. 15a) or the Reynolds number (Fig. 15b) above and below their optimal counterparts. Increasing the Bingham number increases the compression rate and results is a lower final sample height. Increasing the Reynolds number may be reducing the compression rate, but the sample is compressed more and the final sample height is smaller. These observations were crucial in the selection of the optimal parameter values.

Finally, the effects of the thixotropy parameters are illustrated in Fig. 16. As discussed above, the simplified kinetics of Eq. (22) has been employed. From Fig. 16a, it is deduced that the effects of increasing the build-up parameter $a$ are more visible in the final stage of the compression: compression is slower and the final sample height increases with $a$. In the initial stages, however, compression becomes slightly faster when $a$ is increased. In Fig. 16b, it is observed that when the breakdown parameter $b$ is increased, compression eventually becomes faster, the final sample height is smaller, and the compression curve becomes flat, as in the experiments.

Finally, Fig. 17 shows snapshots of the structure parameter contours during the compression experiment when the optimal values of the parameters are used. Initially, the material is fully structured. Once the (modified) load is applied, structure breakdown occurs at the upper outer part of the sample. As a result, the diameter of the sample is bigger at the top and the sample becomes asymmetric. As the time proceeds, the structure break-down propagates towards the axis of symmetry and the sample tends to become symmetric. The evolution of the power-law index within the sample is illustrated in Fig. 18.

Figure 19 illustrates the evolution of the space-averaged values of the structure parameter and the power law exponent, defined by

$$\bar{\lambda}(t) = \frac{2\pi}{V_0} \int_0^{h(t)} \int_0^{R(t)} \lambda(r,z,t) r dr dz \qquad (23)$$

and

$$\bar{n}(t) = \frac{2\pi}{V_0} \int_0^{h(t)} \int_0^{R(t)} n(r,z,t) r dr dz \qquad (24)$$

where $V_0 = \pi R^2(0) h(0)$. The mean values of the structure and power-law index throughout the squeeze experiment are 0.3512 and 0.6212, respectively. The minimum and final mean values of the power-law exponent are 0.4047 and 0.7392, respectively. The variation of $\bar{\lambda}$ during the squeeze experiment when the power-law exponent is constant is qualitatively the same [24]. In other words, this is not affected by the time-dependence of the power-law exponent and the consistency index.

Using chemical composition and porosity data at the end of the solidification, Torres and Zoqui [26] estimated the density of the A356 alloy at 582 ºC to be 2450kg/m$^3$. The rheological parameters of the material are then readily calculated from the optimal dimensionless numbers determined above. The consistency index is calculated from the Reynolds number defined in Eq. (11) to be $k^*$=230 Kg/m/s$^{0.6}$ ($n = n_{max} = 1.4$), while the yield stress (complete structure) from the Bingham number, defined in Eq. (13), is estimated to be $\tau_0^*$=2.27 kPa. Similarly, for the kinetic parameters we find $a^* = 1.91 \times 10^3$ s$^{-1}$, $b$=0.95, and $c^* = 0.20$ s. Figure 20 illustrates various flow curves from no ($\lambda = 0$) to full ($\lambda = 1$) structure. The equilibrium flow curve, that is the steady-state flow curve is also plotted. This is calculated using the equilibrium structure parameter obtained from Eq. (7):

$$\lambda_e = \frac{1}{1+(b/a^*)\dot{\gamma}_e^* e^{c^*\dot{\gamma}_e^*}} \tag{25}$$

It can be observed that the steady-state flow curve coincides with the full- and no-structure flow curves at low and high shear rates, respectively, and that the transition between these two limiting cases occurs for shear rates in the range from 10 to 100 s$^{-1}$. As noted by de Souza Mendes and Thompson [15], the steady-state (equilibrium) flow curve is not represented by the Herschel-Bulkley constitutive equation. The variation of the yield stress and the power-law exponent with the shear rate is shown in Fig. 21. One observes that the material is shear thickening ($n>1$) at low shear rates when structure still exists and shear-thinning ($n<1$) at high shear rates, when structure is destroyed. This behavior can be attributed to the changes in the material microstructure. At low shear rates the microstructure is dendritic and the material exhibits shear thickening; at high shear rates the microstructure becomes rather spheroidal and the behavior of the material in steady-state is always shear thinning [8].

The non-monotonicity of the steady-state flow curve (Fig. 20) was previously noted by Alexandrou et al. [37] who assumed that only the yield stress varies with the structure parameter. Non-monotonic steady-state flow curves have also been obtained from experiments on (thixotropic) colloidal star polymer suspensions [39]. The phenomenon where the steady-state shear stress falls with increasing shear rate, as in the negative-slope regime of the equilibrium flow curve in Fig. 20, has also been observed in experiments on semi-solid alloy slurries in a concentric circular rheometer [40]. McLelland et al. interpret this anomalous behavior in terms of rapid structural breakdown of solid particle clusters on increasing the shear rate from low initial values [40]. Møller et al. [16] pointed out that the negative-slope branch of the steady-state flow curve leads to instability, since a lower stress can sustain a higher shear rate. The theory for this local instability goes back to the work of Yerushalmi et al. [41]. In the present case, the appearance of the negative-slope branch is a consequence of the destruction of structure (at a fixed value of $\lambda > 0$, the material obeys the (positive-slope) Herschel-Bulkley constitutive equation). As a second remark, when neglecting extremely high (and thus unrealistic) values of the shear rate, the two positive-slop branches of the steady-state flow curve in Fig. 20 do not overlap over any range of the shear stress. As a matter of fact, the shear stress is a single-valued function of the shear rate around the central part of the negative-slope branch of the steady-state flow curve.

From Eq. (6), the value of the structure parameter at which $n=1$ is $\lambda = (1-n_{min})/(n_{max}-n_{min})$. Substituting into Eq. (25) and solving for the shear rate, one finds that the critical shear rate at which $n=1$ at equilibrium is given by

$$\dot{\gamma}_1^* = \frac{1}{c^*} W\left(\frac{a^* c^* (n_{max}-1)}{b(1-n_{min})}\right) \approx 19.67 \text{ s}^{-1} \qquad (26)$$

where $W$ is the principal branch of Lambert's W function [28,42].

## 5. Conclusions

A method for determining the time-dependent material parameters of semisolid metals from squeeze flow data has been proposed. The Herschel-Bulkley constitutive equation has been employed along with a structural thixotropic model and finite elements have been used in a Lagrangian framework in order to reproduce squeeze constant-load experimental data on a semisolid aluminium alloy A356. Our systematic runs for wide ranges of the material and flow parameters showed that excellent agreement with the experimental data could be achieved provided that at the initial stages of the experiment, i.e., for times up to 0.01 s in the 0.9 s experiment, the applied load is 10 times higher than the nominal experimental load and that both the yield stress and the power-law exponent vary linearly with the structure parameter. The first assumption implies that a different model, such as an elastoviscoplastic one, needs to be employed during the initial stages of the experiment. As for the second one, the evolution of the sample height can be reproduced allowing the yield stress to vary from 0 (no structure) to a maximum nominal value (full structure) and the power-law exponent from 0.2 to 1.4, i.e., from the shear-thinning to the shear-thickening regime. These variations are consistent with the internal microstructure variations known to be exhibited by semisolid slurries and the arguments of Atkinson and Favier [8], who analyzed systematically literature data to rationalize apparent contradictions regarding shear thickening in semisolid materials. At low shear rates, the dendritic microstructure prevails and the material is shear thickening, whereas at higher shear rates microstructure becomes spheroidal and the material is shear thinning. Atkinson and Favier [8] argue that if the semisolid suspension is dis-agglomerated before the shear rate jump, the instantaneous behavior upon a sudden shear rate increase is shear thickening provided the solid fraction is greater than 0.36 and the final shear rate is greater than about 100 s$^{-1}$. They also note that shear thickening may be masked by yield stress if the sudden shear-rate increase is from rest.

Estimates of the material parameters at 582 °C have been obtained using the optimal values of the various dimensionless parameters. The determination of the material constants could be a useful tool in optimizing the semi-solid process.

Computational rheology not only enables the rheological characterization of thixotropic semisolid metal slurries, but also sheds light on the flow characteristics and the microstructure evolution. The numerical results show that thixotropy does not affect the flow initially. However, once the structure is destroyed, the unyielded regions grow slower than in the non-thixotropic case allowing for longer compression of the sample. Under constant force, the structure may be destroyed in the early stages of the compression, but it then re-builds steadily till the cessation of the flow experiment. As for the future work, we will consider the use of an elasto-thixo-viscoplastic model in order to describe well the initial stages of the compression experiment. We are plan to investigate possible wall slip effects.

## Declaration of competing interest

The authors declare that they have no known competing financial interests or personal relationships that could have appeared to influence the work reported in this paper.

## Data availability

Data will be made available on request.

## Author contributions

The present study was conceptualized by G.C. Florides, G.C. Georgiou, M. Modigell, and E.J. Zoqui. G.C. Florides performed the simulations, analyzed and visualized the data, and drafted the manuscript. E.J. Zoqui carried out the experiments and draft the experiments section of the paper. The original manuscript draft was extensively reviewed and edited by G.C. Georgiou, M. Modigell, and E.J. Zoqui. The research project was supervised by G.C. Georgiou and M. Modigell.

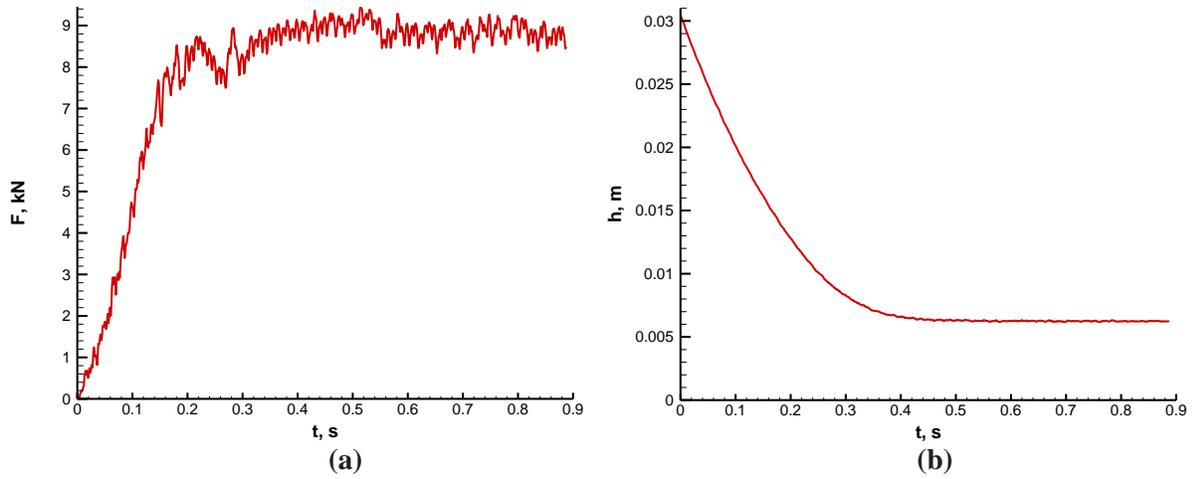

**Figure 1**. Experimental data on an A356 alloy at 582 °C (semisolid stage): (a) applied load; (b) evolution of the height.

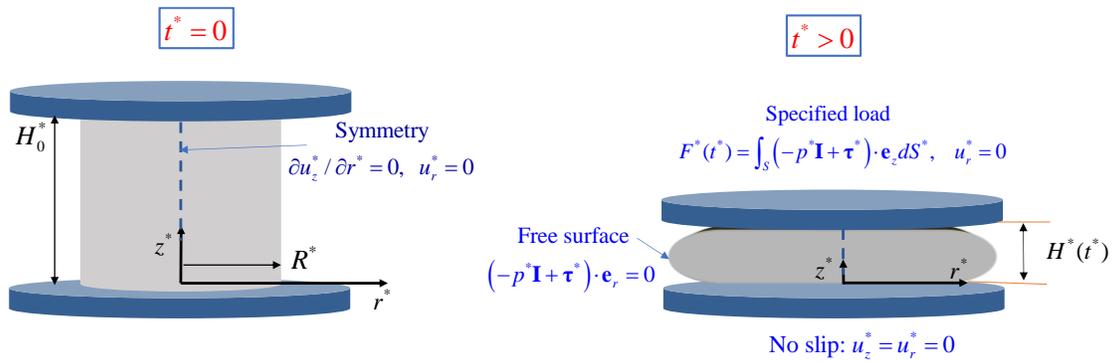

**Figure 2**. Geometry and boundary conditions of squeeze flow when a load is applied on the top plate.

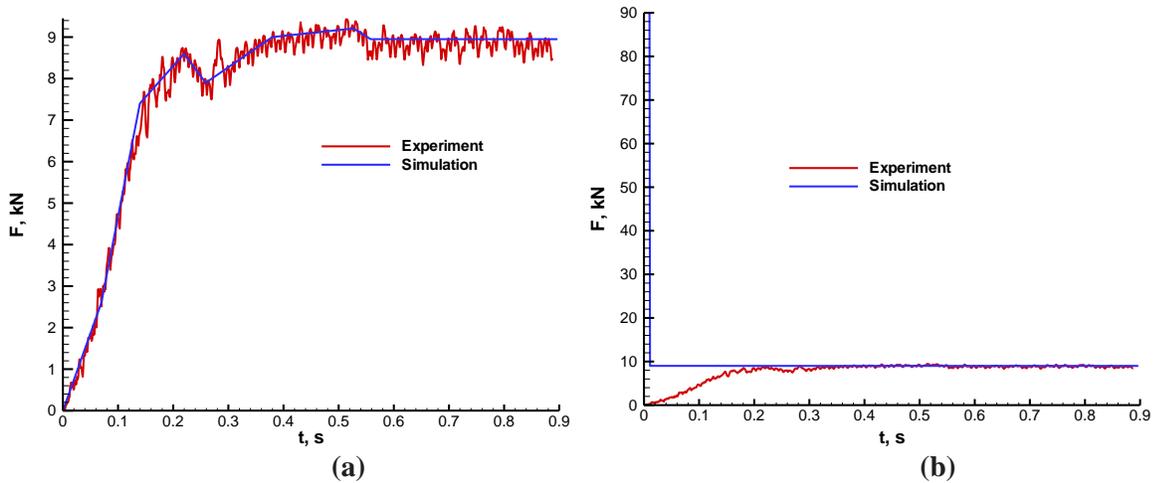

**Figure 3**. Load distributions employed in the numerical simulations: (a) Distribution following the experimental data; (b) Two-step load with a high value for the first 0.01 s and a lower constant value afterwards equal to nominal maximum experimental load.

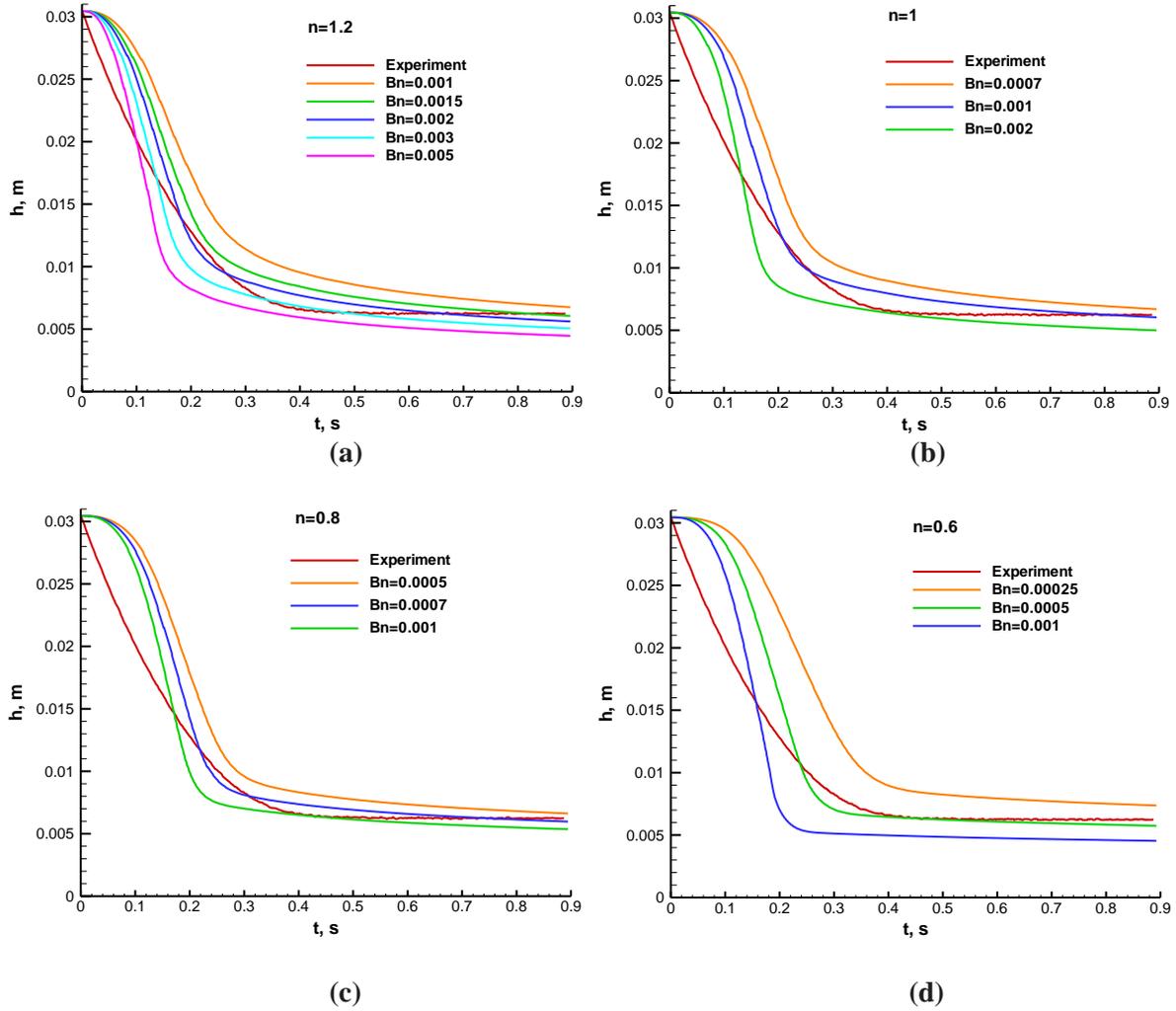

**Figure 4**. Effect of the Bingham number for different power-law indices when $Re=1$ and $a=b=1$: (a) $n=1.2$; (b) $n=1$; (c) $n=0.8$; (d) $n=0.6$; The experimental load distribution is imposed.

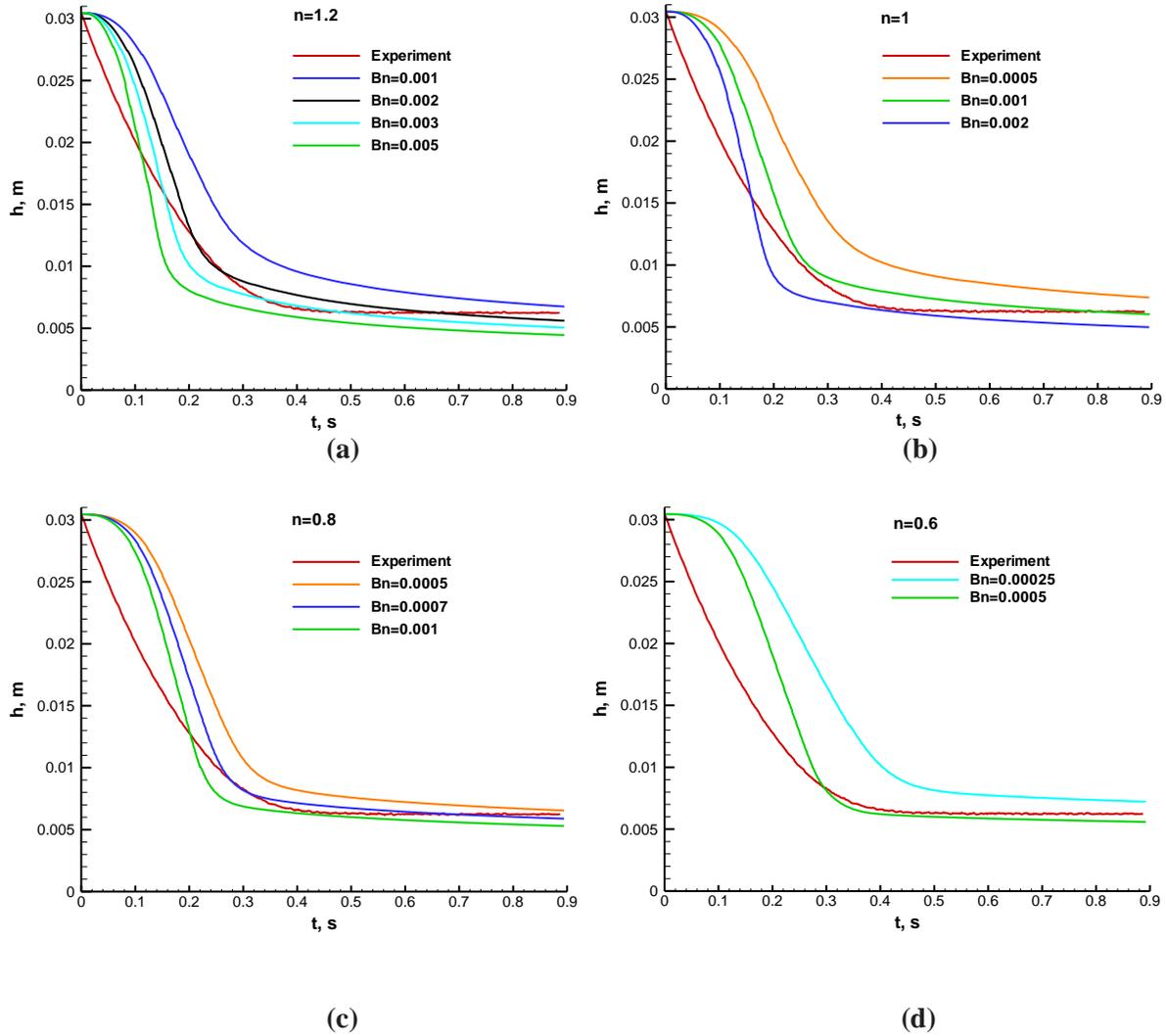

**Figure 5**. Effect of the Bingham number for different power-law indices when $Re = 1.5$ and $a = b = 1$: (a) $n = 1.2$; (b) $n = 1$; (c) $n = 0.8$; (d) $n = 0.6$; The experimental load distribution is imposed.

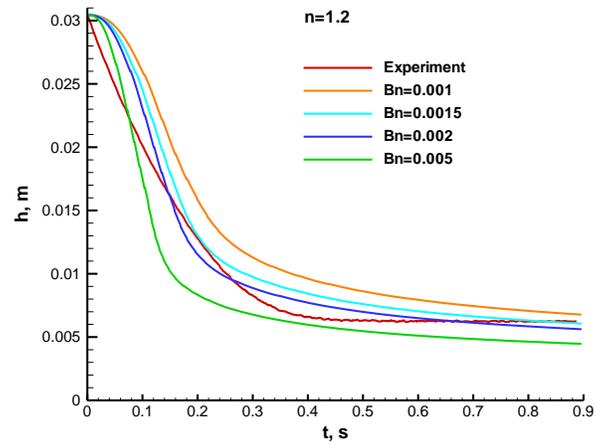

(a)

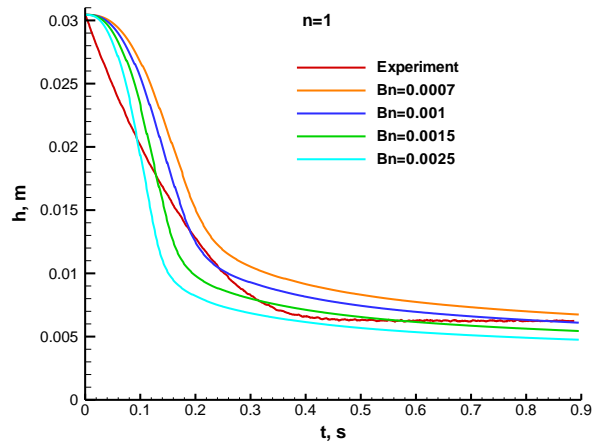

(b)

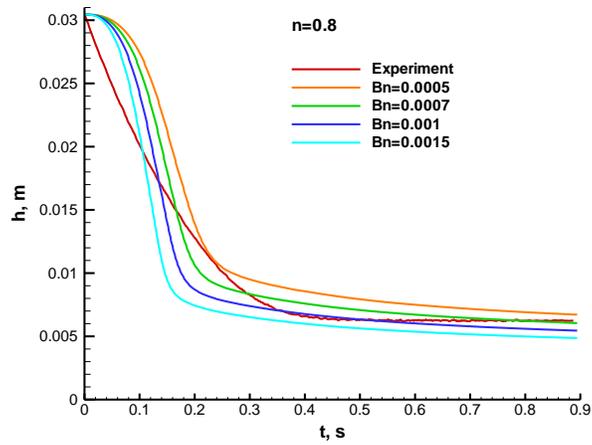

(c)

**Figure 6**. Effect of the Bingham number for different power-law indices when $Re = 0.5$ and $a = b = 1$: (a) $n = 1.2$; (b) $n = 1$; (c) $n = 0.8$; The experimental load distribution is imposed.

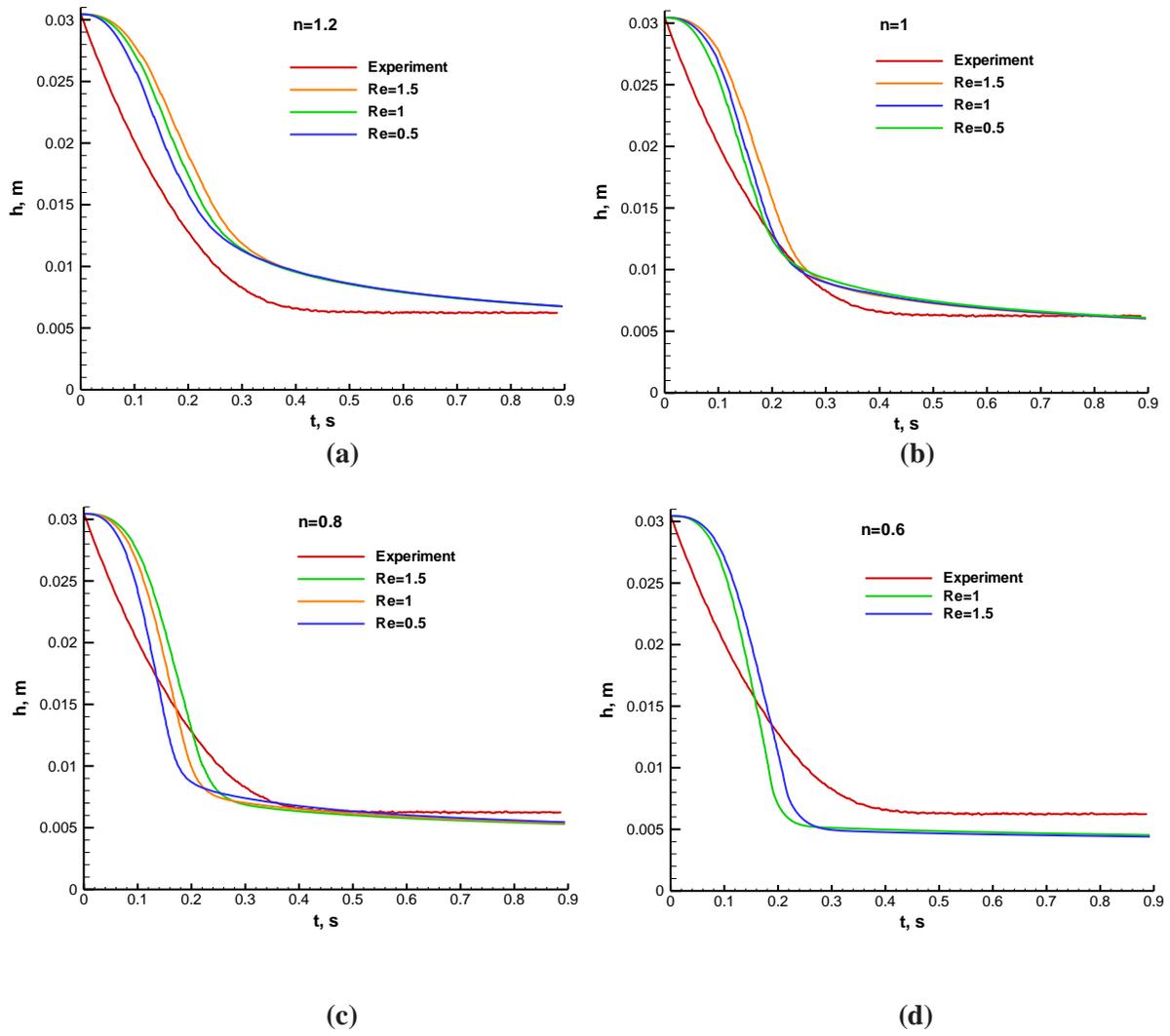

**Figure 7**. Effect of the Reynolds number for different power-law indices when $Bn = 0.001$ and $a = b = 1$ : (a) $n = 1.2$; (b) $n = 1$; (c) $n = 0.8$; (d) $n = 0.6$; The experimental load distribution is imposed.

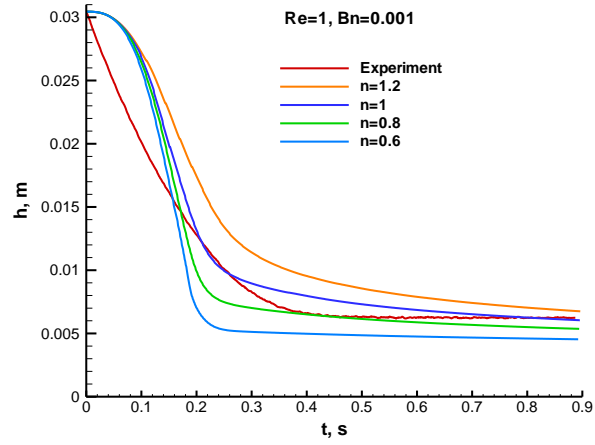

(a)

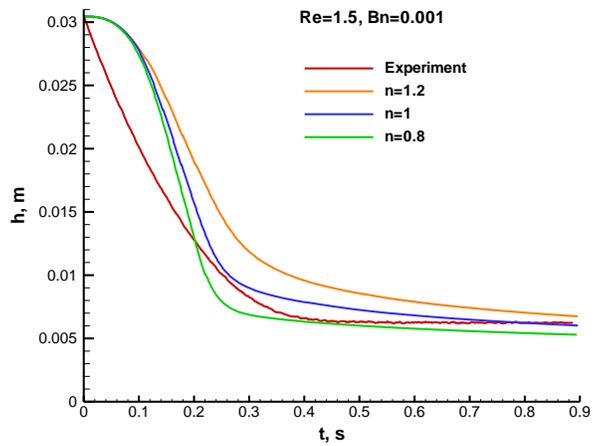

(b)

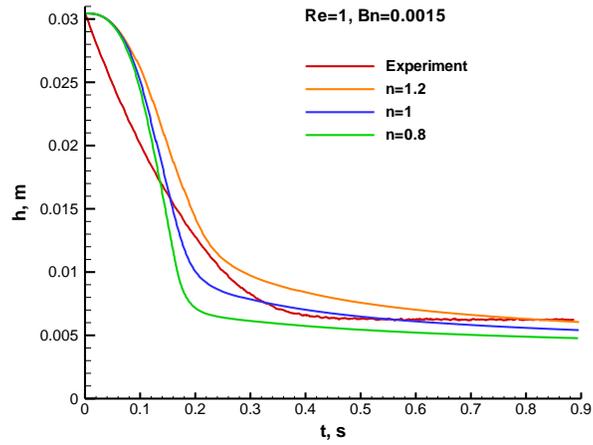

(c)

**Figure 8**. Effect of the power-law index when $a = b = 1$: (a) $Re = 1, Bn = 0.001$; (b) $Re = 1.5, Bn = 0.001$; (c) $Re = 1, Bn = 0.0015$; The experimental load distribution is imposed.

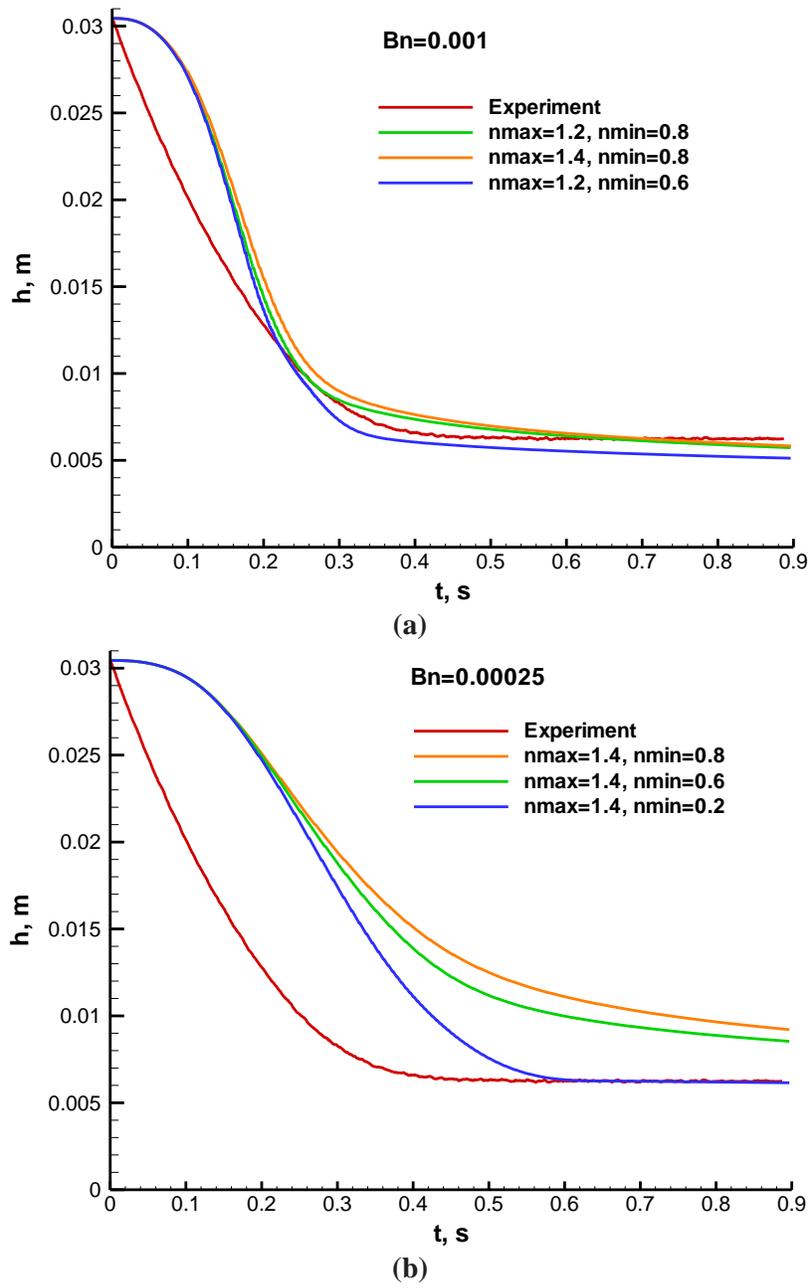

**Figure 9**. Results with variable power-law index for $Re=1$ and $a=b=1$: (a) $Bn=0.001$; (b) $Bn=0.00025$; The experimental load distribution is imposed.

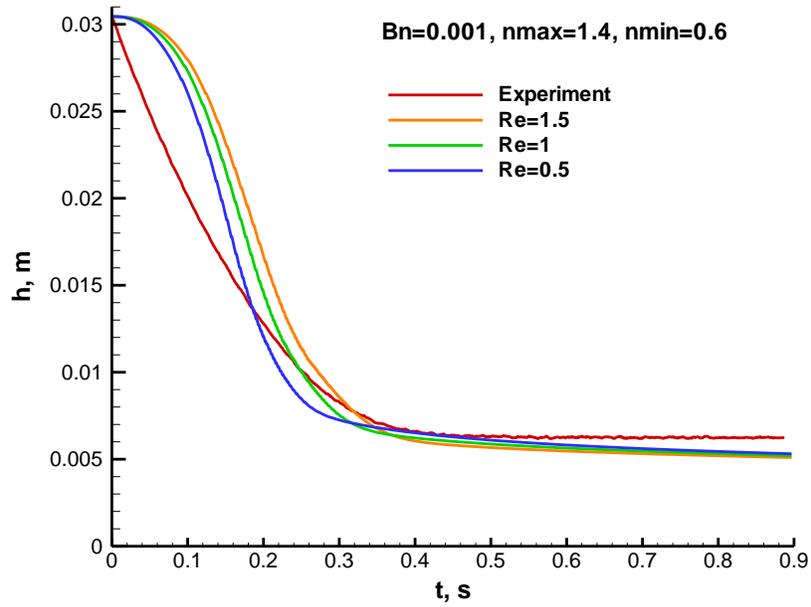

**Figure 10**. Results for $Bn = 0.001$, $a = b = 1$ and various Reynolds numbers when the power-law index varies from 0.6 to 1.4. The experimental load distribution is imposed.

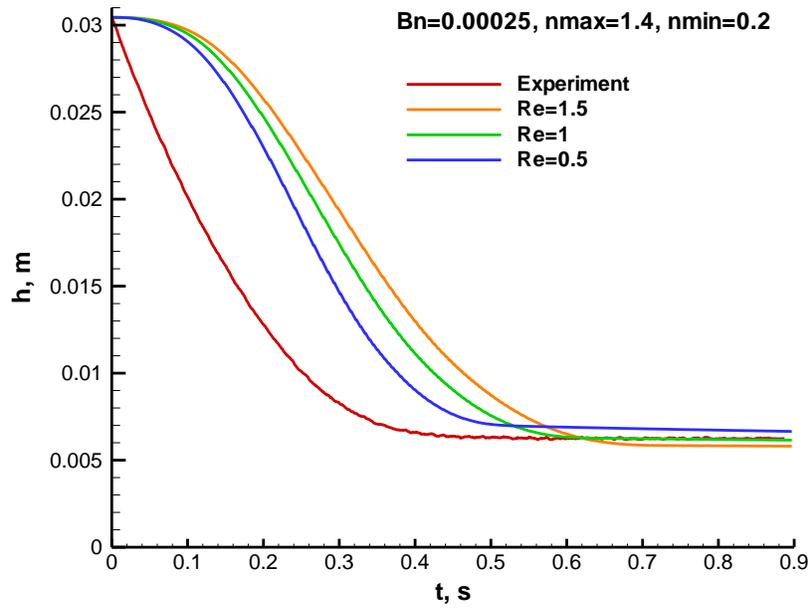

**Figure 11**. Results for $Bn = 0.00025$, $a = b = 1$ and various Reynolds numbers when the power-law index varies from 0.2 to 1.4. The experimental load distribution is imposed.

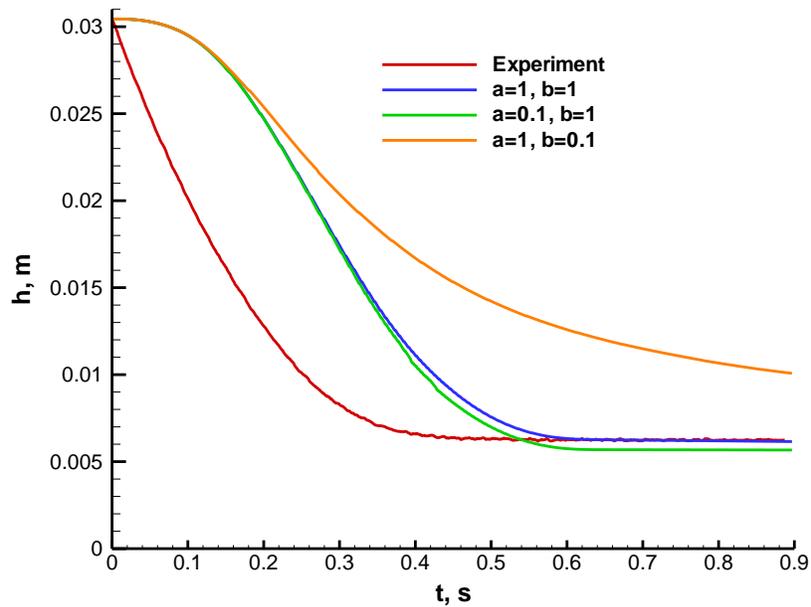

**Figure 12**. Effect of the kinetic parameters $a$ and $b$ when $Re=1$, $Bn=0.00025$, and the power-law index varies from 0.2 to 1.4. The experimental load distribution is imposed.

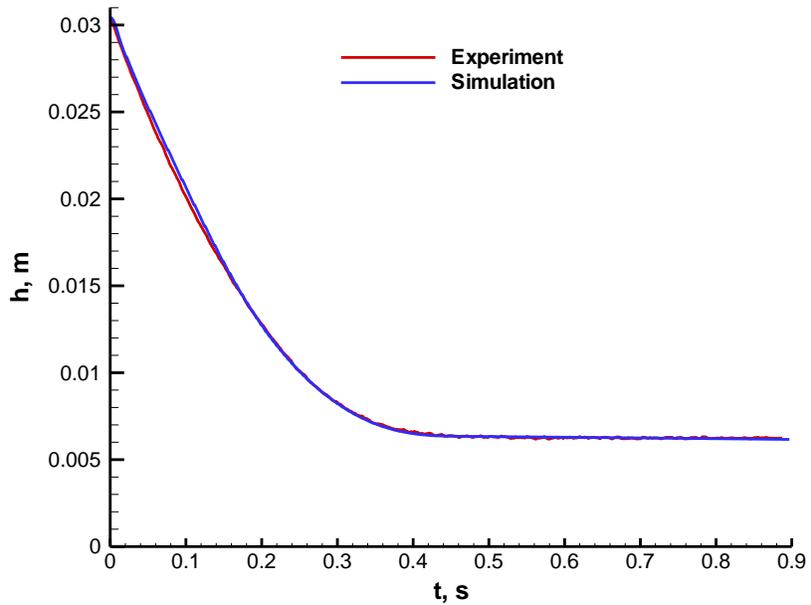

**Figure 13**. Experimental data are reproduced when the load in the initial 0.01s of the experiment is 10 times the nominal maximum experimental load $F_\infty^*=9$ kN and the power-law index varies from 0.2 to 1.4. The optimal values of the other parameters are $Re=0.95$, $Bn=0.000235$, and $a=b=0.95$.

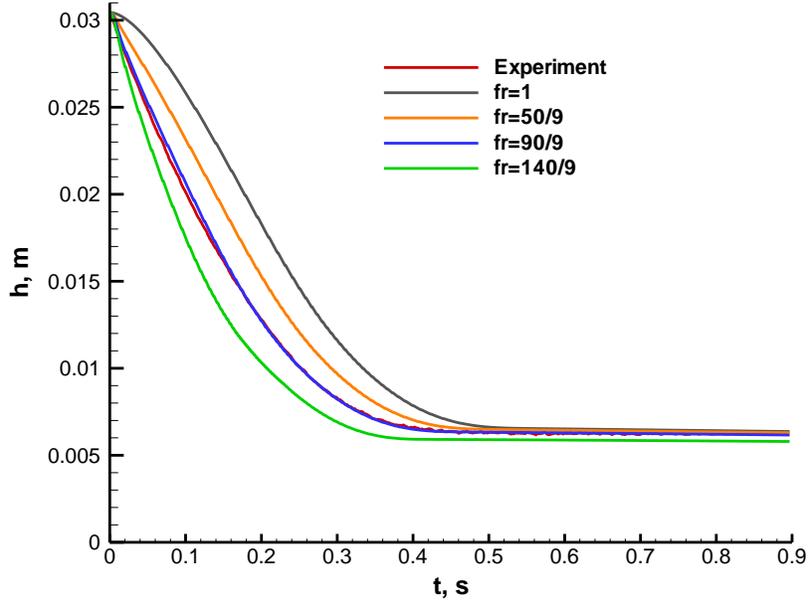

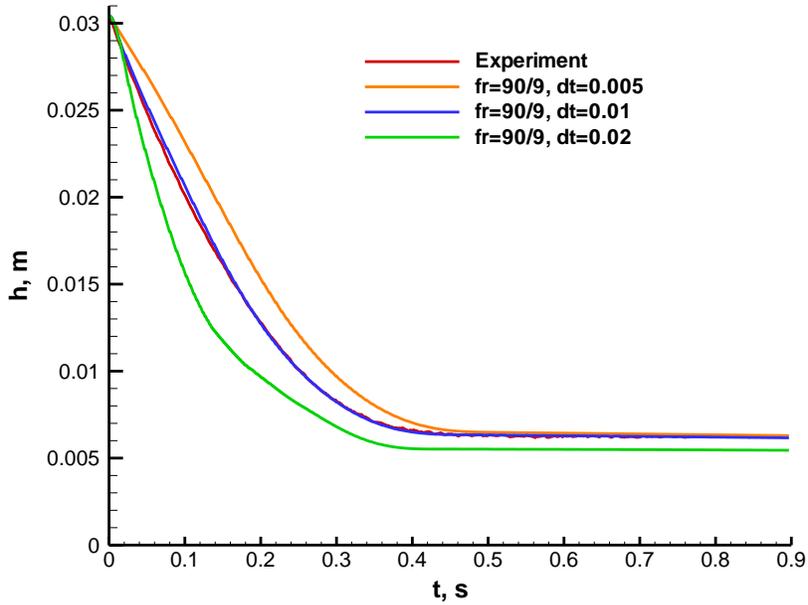

**Figure 14**. Effects of (a) the load ratio when $\Delta t^* = 0.01$ s and (b) $\Delta t^*$ when the load ratio is 10 and the other parameters are optimal: $Re = 0.95$, $Bn = 0.000235$, $a = b = 0.95$ and the power-law index varies from 0.2 to 1.4.

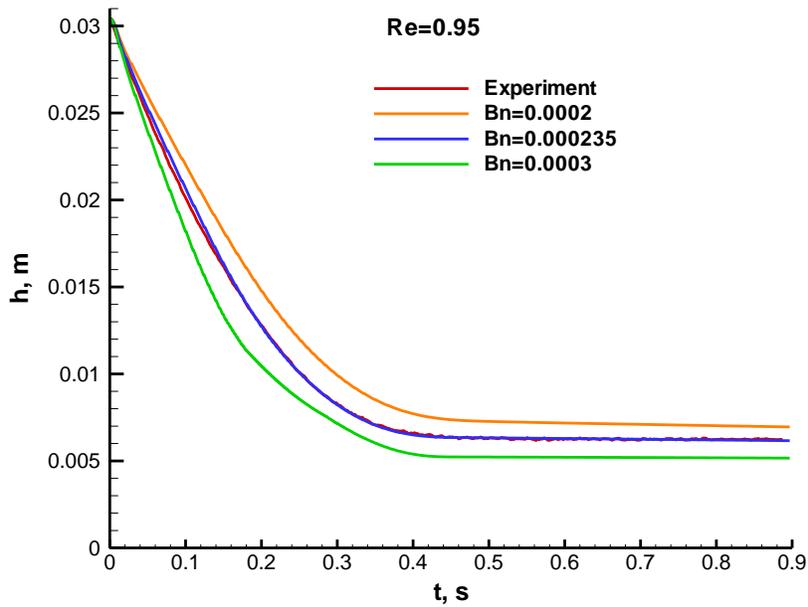

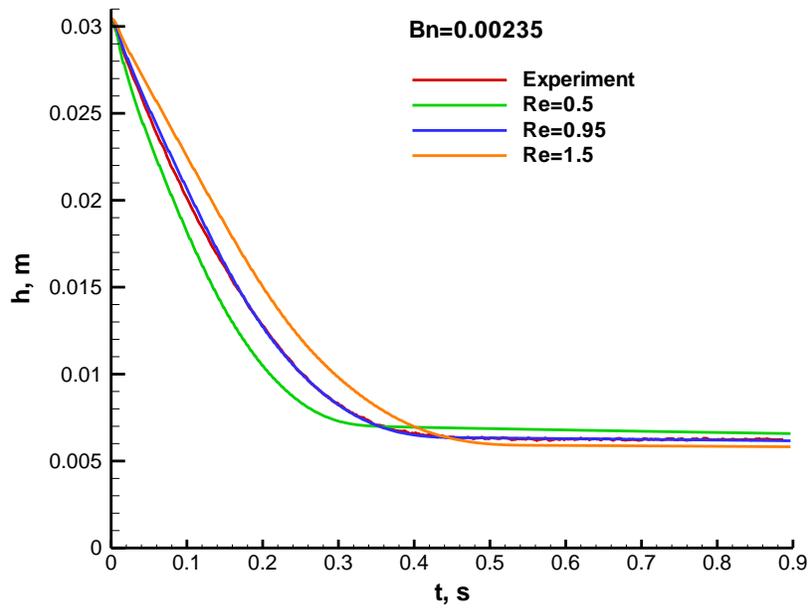

**Figure 15.** Effects of (a) the Bingham number when $Re = 0.95$ and (b) the Reynolds number when $Bn = 0.000235$ and the other parameters are optimal: $a = b = 0.95$, the initial load for the first 0.01 s is 90 kN and the power-law index varies from 0.2 to 1.4.

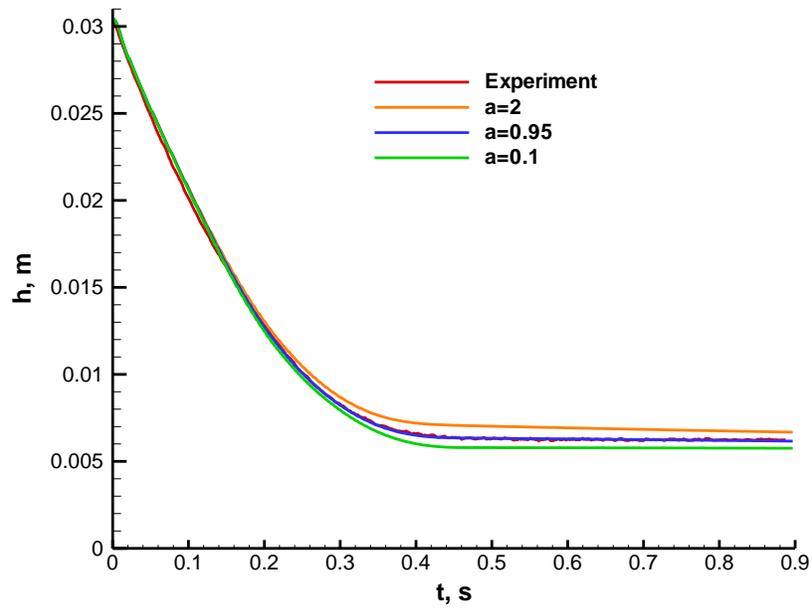

(a)

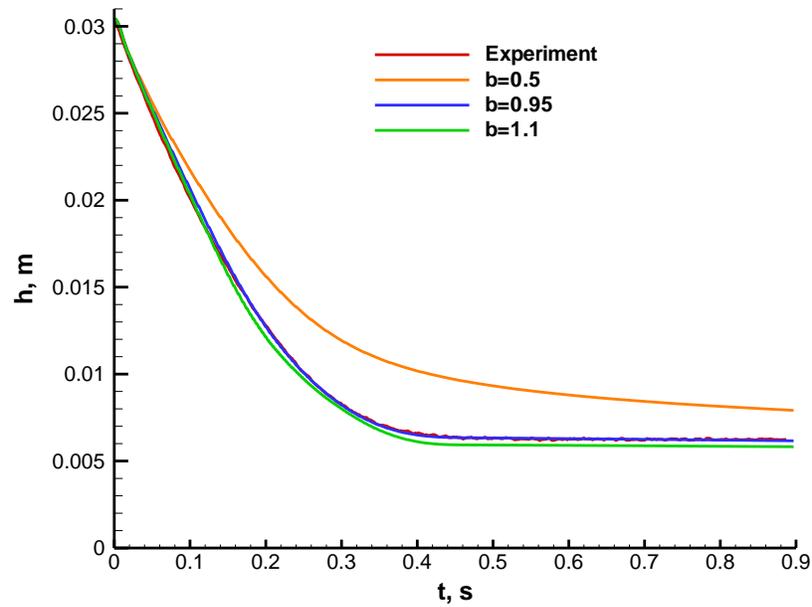

(b)

**Figure 16**. Effects of (a) *a* when $b = 0.95$ and (b) *b* when $a = 0.95$ and the other parameters are optimal: $Re = 0.95$, $Bn = 0.000235$, the power-law index varies from 0.2 to 1.4, and the initial load for the first 0.01s is 9 kN.

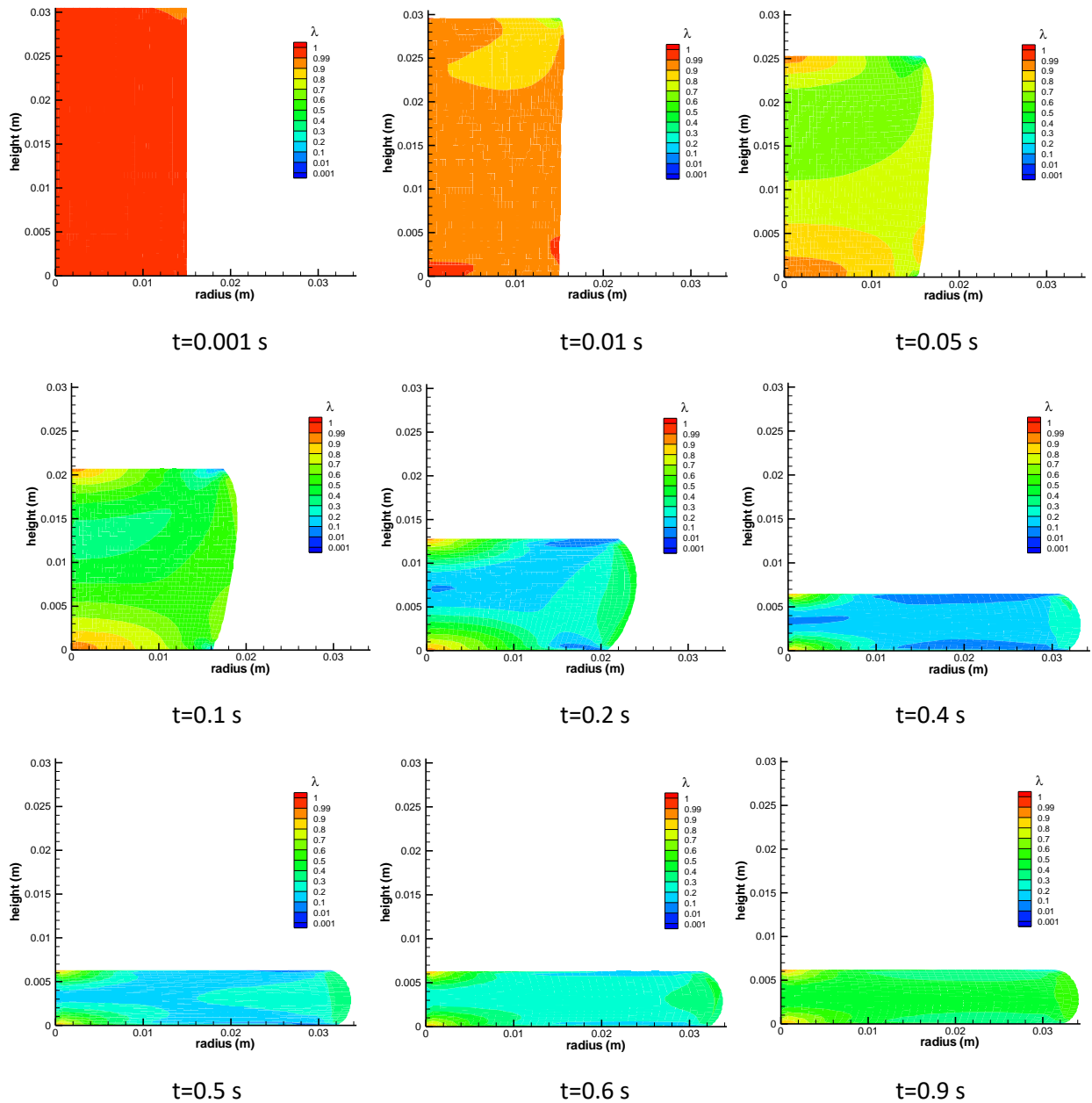

**Figure 17**. Evolution of the structure parameter during compression using the optimal parameters. $Re = 0.95$, $Bn = 0.000235$, the power-law index varies from 0.2 to 1.4, and the initial load for the first 0.01s is 90 kN.

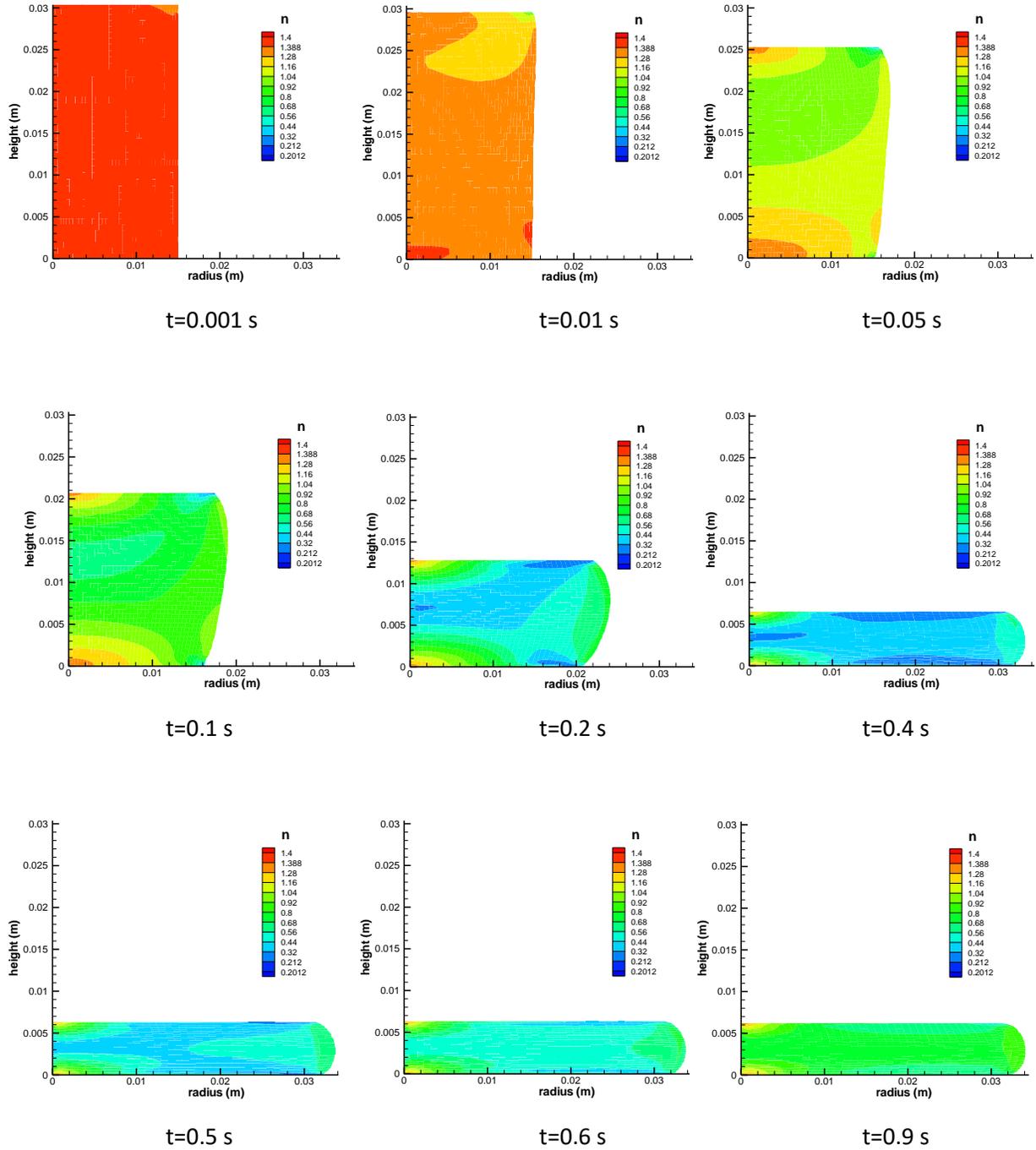

**Figure 18**. Evolution of the power law index *n* during compression using the optimal parameters. $Re = 0.95$, $Bn = 0.000235$, the power-law index varies from 0.2 to 1.4, and the initial load for the first 0.01s is 90 kN.

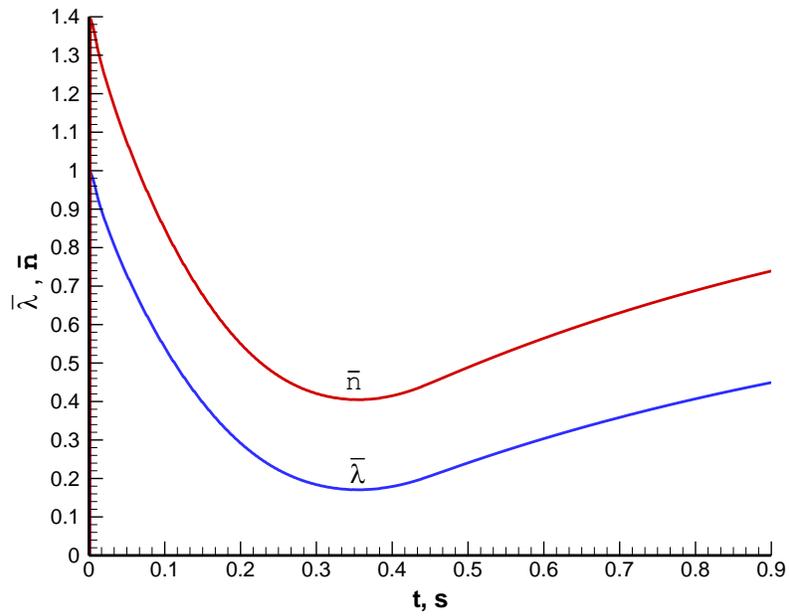

**Figure 19**. Evolutions of the mean values of the structure parameter ($\bar{\lambda}$) and the power law index ($\bar{n}$) during compression using the optimal parameters. $Re = 0.95$, $Bn = 0.000235$, the power-law index varies from 0.2 to 1.4, and the initial load for the first 0.01s is 90 kN.

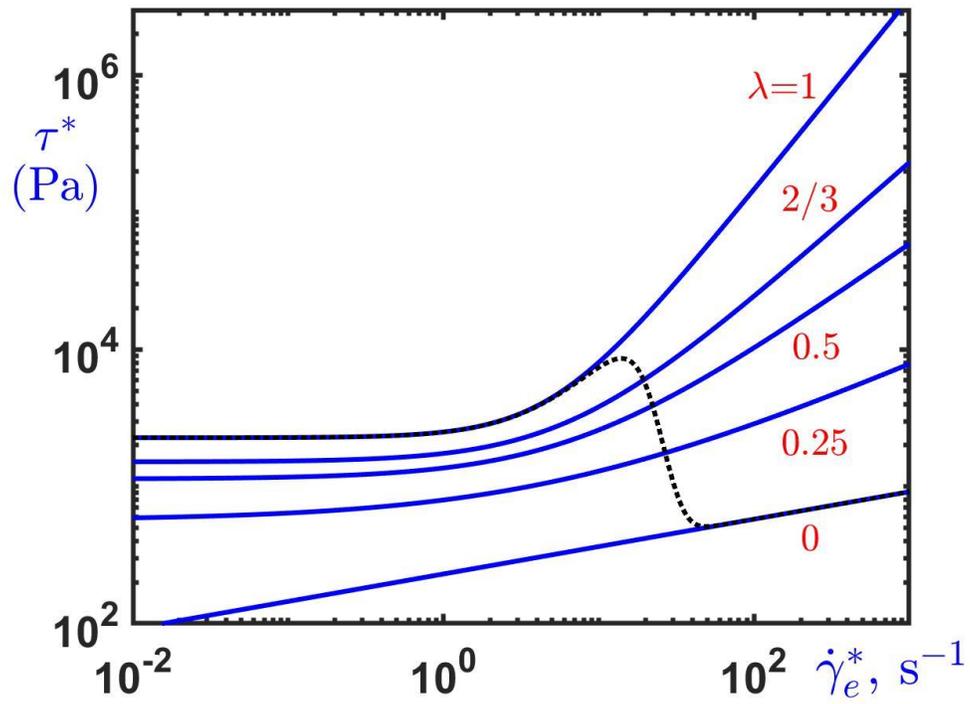

**Figure 20**. Constant structure flow curves for $\lambda = 0$ (no structure), 0.25, 0.5, 2/3 and 1 (full structure) and equilibrium flow curve (dotted line) with the estimated material parameters: $\tau_0^* $=2.27 kPa, $k^*$ =230 Kg/m/s$^{0.6}$, $n_{min} = 0.2$, $n_{max} = 1.4$, $a^* = 1.91 \times 10^3$ s$^{-1}$, $b$=0.95, and $c^* = 0.20$ s.

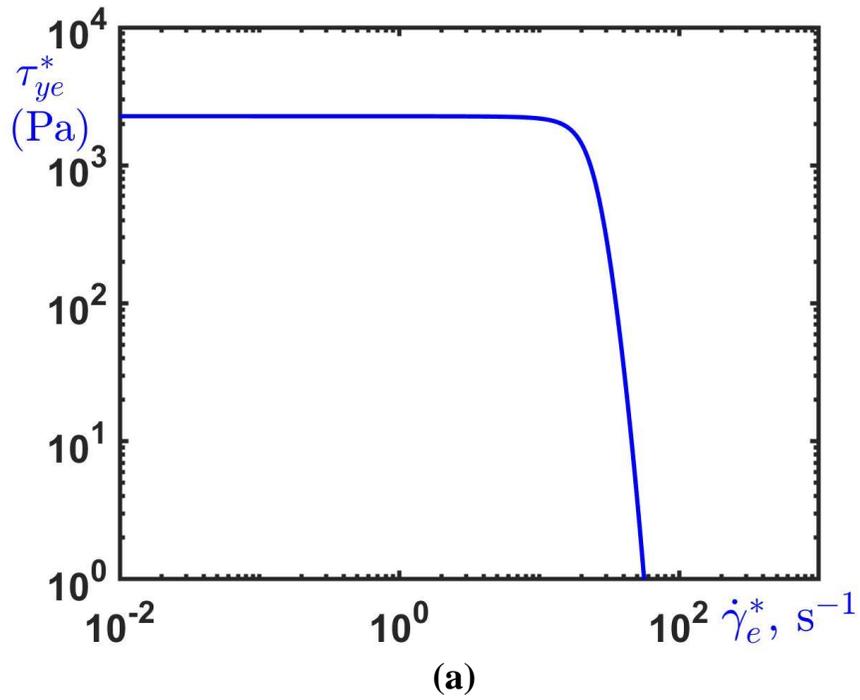

(a)

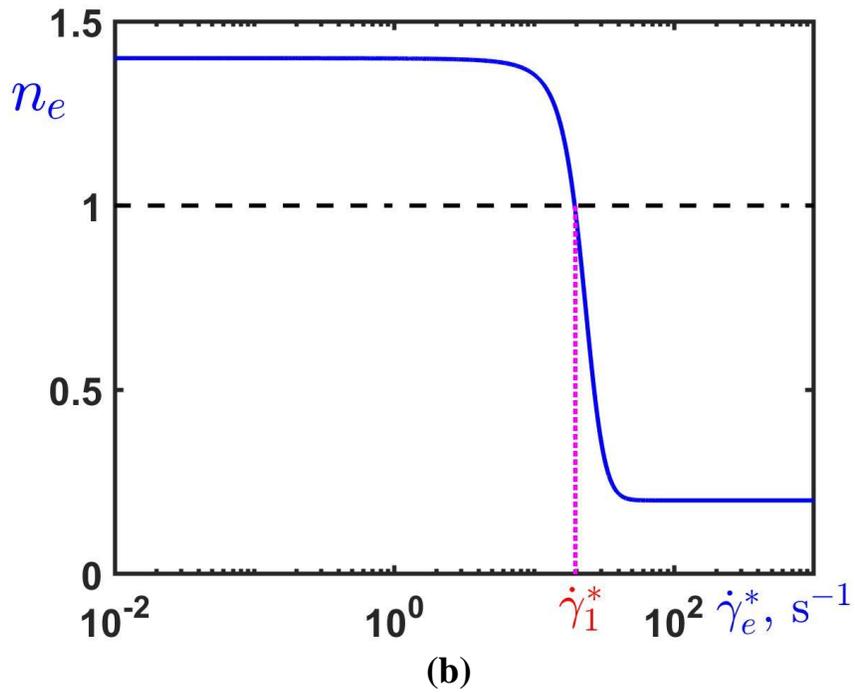

(b)

**Figure 21**. Variation of (a) the yield stress and (b) the power-law exponent with the shear rate at equilibrium calculated using the estimated material parameters, i.e., $\tau_0^*$=2.27 kPa, $k^*$=230 Kg/m/s$^{0.6}$, $n_{min} = 0.2$, $n_{max} = 1.4$, $a^* = 1.91 \times 10^3$ s$^{-1}$, $b$=0.95, and $c^* = 0.20$ s.